\documentclass[aps,prl,twocolumn,groupedaddress,epsfig]{revtex4}
\usepackage{graphicx,amssymb}

\usepackage{amssymb} 

\begin{document}
\title{Protein crystallization in confined geometries}
\author{S. Tanaka, S. U. Egelhaaf and W. C. K. Poon}
\affiliation{School of Physics and Astronomy, The University of Edinburgh, JCMB, Kings Buildings, Mayfield Road, Edinburgh EH9 3JZ, UK}    

\begin{abstract}   
We studied the crystallization of a globular protein, lysozyme, in the cubic phase of the lipid monoolein. The solubility of lysozyme in salt solution decreased by a factor of $\sim 4$ when confined in cubic phase. Calculations and Monte Carlo simulations show that this can be explained by the {\it confinement} of lysozyme molecules to the narrow water cells in the cubic phase. 
\end{abstract} 


\maketitle

We report a study of the crystallization of lysozyme, a globular protein, in the cubic phase of monoolein, an uncharged lipid. The protein solubility was found to decrease by a factor of $\sim 4$. We account for this observation by the {\it confinement} of lysozyme molecules within nanoscopic water `cells' in the cubic phase. The resulting entropy-driven rise in their chemical potential can explain our observation of solubility suppression compared to the lipid-free case. The generic nature of this explanation means that our results are of potential interest to three rather different audiences. First, and most generally, scientists interested in the basic physics of confinement \cite{Gelb99} may find the lysozyme-monoolein system a useful model for detailed study. Secondly, there is growing interest in directing the self-assembly of synthetic or natural nanoparticles using various means of confinement (e.g. at an interface \cite{Dinsmore03} or in emulsion droplets \cite{Pine03}). Our work suggests that lipid cubic phases can be useful in this regard. Thirdly, and most specifically, lipid cubic phases offer an environment for {\it globular} protein crystallization \cite{McPherson99}. (Their use for crystallizing membrane proteins \cite{Landau96}, which reside in the lipid bilayers, relies on different physics \cite{Caffrey00,grabe03}.)  

Lysozyme crystallization in lipid cubic phase has been reported before \cite{Landau97}, although to date there is no complete phase diagram or physical explanation of the mechanism. We have studied the phase behavior of lysozyme solution as a function of protein and salt concentration in a monoolein cubic phase, and compared it with the case {\it without} lipids. The cubic phase was found to lower the solubility dramatically. This result was then explored using simple calculations and Monte Carlo simulations.

Lysozyme and monoolein (MO) were used as purchased from Sigma-Aldrich. Protein concentration in 0.05~M sodium acetate buffer was determined, after filtration through a 0.1 $\mu$m Millipore filter, by UV spectrophotometry (specific absorbance $A_{\rm 280nm} = 2.64$~ml/mg.cm). NaCl solutions of known concentration were then added. 

\begin{figure}
\begin{center}
\rotatebox{0}{\includegraphics[width=5cm]{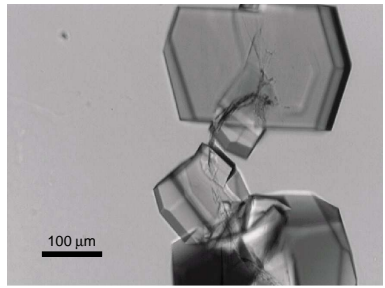}}
\caption{An optical micrograph of lysozyme crystals in monoolein cubic phase. Note the curved piece of solid inclusion; other crystals grow next to surfaces.}
\label{micrograph}
\end{center}
\end{figure}

We first determined the bulk solubility boundary of lysozyme/salt without lipid from the dissolution of `seed' crystals \cite{McPherson99}, which occurs at lower protein and salt concentrations than the bulk nucleation boundary (where crystals are first seen) \cite{pH4.5phase} due to the nucleation barrier \cite{McPherson99}. Next protein/salt solutions with concentrations {\it below} the bulk solubility boundary was added to and vigorously mixed with MO powder melted on a microscope slide. A highly viscous cubic phase was observed to form almost instantly. The excess solution was then removed and the cubic phase sealed under a cover slip (sample thickness $\sim 100$ $\mu$m) for observation in an optical microscope at 18$^{\circ}$C. All samples were transparent and optically isotropic under cross polarizers. Within two months, crystals were observed in many samples, e.g. Fig.~\ref{micrograph}.

\begin{figure}
\begin{center}
\rotatebox{-90}{\includegraphics[width=7cm]{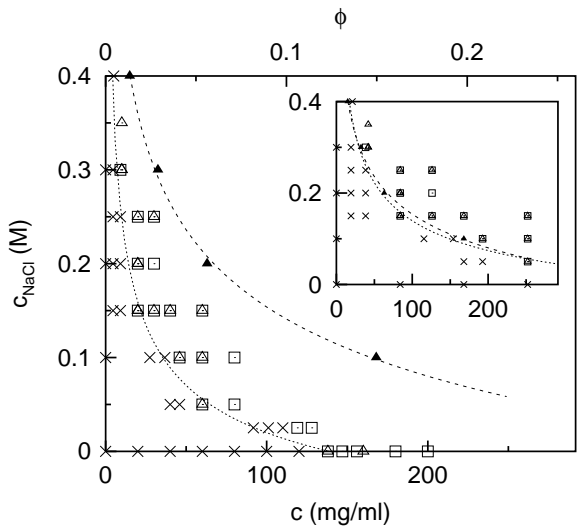}}
\caption{The phase diagram of lysozyme/NaCl in bulk solution and in MO cubic phase. Filled triangles ($\blacktriangle$) represent  the solubility limit in bulk solution. Other symbols represent behavior in the cubic phase: homogeneous solution ($\times$),  crystals ($\square$), and other precipitates ($\triangle$). The lines are guides to the eye. Inset (same axes), the concentration of the samples in the cubic phase is multiplied by 4, to nearly superimpose the nucleation boundary {\it in cubo} on the bulk solubility boundary.}
\label{phase_diagram}
\end{center}
\end{figure}

Fig.~\ref{phase_diagram} summarizes our observations in the form of a phase diagram. We give lysozyme concentrations, $c$, in mg/ml, and in terms of an effective volume fraction, $\phi$, by modelling the globular protein as a sphere of diameter 3.4~nm (which reproduces the molecular volume). Note that for the data points recording behavior {\it in cubo}, the protein/salt concentrations given are those of the parent solution into which MO powder was added. The boundary above which we observed crystals {\it in cubo} (a kinetic nucleation boundary) lies entirely below the bulk solubility boundary. Since we expect the {\it in cubo} solubility boundary to lie at even lower protein/salt concentrations, our main experimental finding is that lipid cubic phase significantly depresses lysozyme solubility. 

\begin{figure}
\begin{center}
\includegraphics[width=3.6cm]{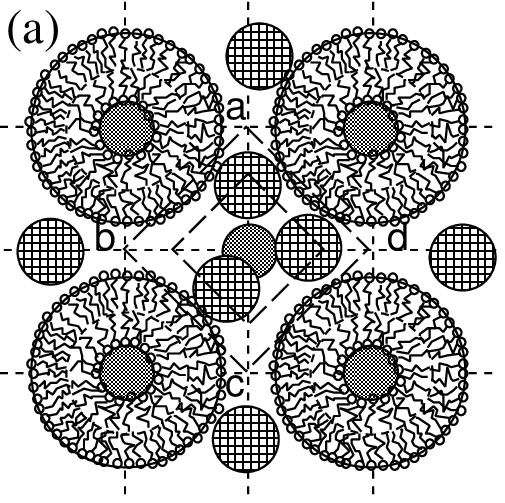}
\includegraphics[width=3.6cm]{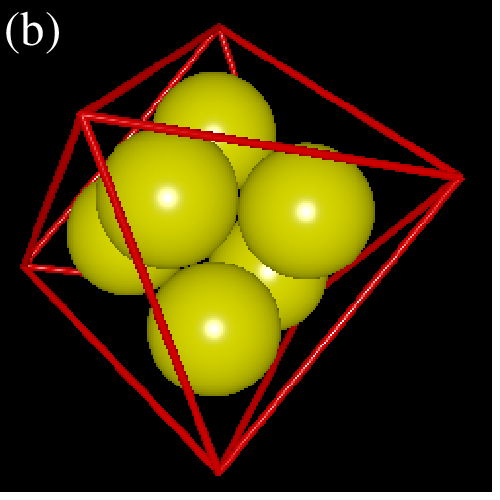}
\caption{(a) A 2-d projection of the {\it Im3m} cubic phase, with `necks' ($a$-$d$) connecting water `cells'. Annular regions are lipid bilayers (with lipids shown schematically). Shaded circles are cross sections of `necks' in and out of the plane of the paper. Hatched particles are (arbitrarily-placed) lysozymes with diameter $\sigma=3.4$~nm. The lattice parameter is 12.5~nm. The larger square (broken line) is the base of a regular square bipyramid used to model a water `cell'. The smaller square is the base of the (smaller) bipyramid accessible to the centres of lysozyme molecules. (b) A bipyramid `cell' with lysozyme molecules.}  
\label{Im3m}
\end{center}
\end{figure}

To understand this observation, we first note that the bicontinuous water in the cubic phase is made up of cells connected by narrow `necks'.  The presence of lysozyme changes the MO cubic phase structure from {\it Pn3m} to {\it Im3m} \cite{Razumas96}, a schematic projection of which is shown in Fig.~\ref{Im3m}(a) with an inter-cell distance of 12.5~nm \cite{Razumas96} and bilayer thickness of 3.5~nm \cite{Caffrey00}. We have also shown a number of lysozyme molecules (hatched circles) modelled as spheres with diameter $\sigma=3.4$ nm. Since the bending rigidity of the bilayers is $k_c \approx 9 k_B T$ at room temperature \cite{Chung94}, proteins cannot move between cells without incurring significant cost in distorting the connecting `necks'. Thus the lysozyme will largely be confined in nanoscopic water `cells' with linear dimension $\lesssim 10$~nm, each of which can hold up to 7 lysozyme molecules \cite{Razumas96,structures}. The center of a lysozyme molecule does not come closer than $\sim \sigma/2 = 1.7$~nm to a bilayer. While such excluded volume is a small effect in the bulk, it is highly significant inside a water `cell' in the cubic phase, leading to a large increase in the effective protein concentration. This explains why crystallization occurred at lower {\it average} protein concentration.

Enquiring {\it where} the crystals actually nucleate gives further understanding. The critical nucleus in our experiments is expected to consist of $> 10$ lysozymes \cite{Galkin00}, more than an undistorted water `cell' can hold. Moreover, a (tetragonal) unit cell contains 8 molecules \cite{Blake65}, and it is unlikely that crystals can nucleate in pores smaller than a few unit cells \cite{Mu96}. Experimentally, we often observed crystals next to solid inclusions or next to surfaces, Fig.~\ref{micrograph}. We therefore propose a role for {\it defects}.

The picture is as follows. The protein in bulk solution has a concentration-dependent chemical potential $\mu(c)$. Confinement {\it in cubo} increases the effective lysozyme concentration to $c^{\prime} > c$, and its chemical potential to $\mu(c^{\prime}) > \mu(c)$. There are defects in the cubic phase of mesoscopic sizes ($\gg$ lysozyme diameters), due to grain boundaries, distortion by inclusions, or surfaces. In these defects, the chemical potential of lysozyme is initially at $\mu(c)$, but diffusion from the `loaded' cubic phase rapidly increases this to close to $\mu(c^{\prime})$. Crystals nucleate from this concentrated solution residing in the defects. Subsequently, as a crystal grows, it will break the surrounding matrix; the elasticity of the cubic phase may therefore affect the rate of crystal growth. 

Support for the above picture comes from measuring the time-dependent protein concentration in lysozyme/salt solution in contact with the cubic phase formed by mixing it with MO.  We observed  (data not shown) a rapid increase in concentration that continued for many hours, showing that indeed $\mu(c^{\prime}) > \mu(c)$.  

To explore quantitatively our proposed picture, we used Monte Carlo (MC) simulations to study attractive hard spheres confined within a `model cubic phase'. To do this, we first need a numerical estimate of the excluded volume effect. The center of a lysozyme molecule confined to a water `cell' in the cubic phase cannot explore the whole of the cell volume, $V_c$, but only the accessible volume, $V_a$, given by taking off the volume of a layer of thickness $\sigma/2$ from $V_c$. The ratio $\alpha = V_c/V_a$ characterizes the excluded-volume effect.

Estimating $\alpha$ for the $Im3m$ structure involves non-trivial geometry and knowing the exact lattice parameter, $a$, pertaining to our experiments. The value of $a$ used in Fig.~\ref{Im3m} is for fully-hydrated MO cubic phase equilibrated at 25$^{\circ}$C \cite{Razumas96}. These conditions almost certainly do not apply (after mixing at 25$^{\circ}$C we immediately removed excess solution and observed at 18$^{\circ}$C). Nevertheless, using $a = 12.5$~nm \cite{Razumas96} as an estimate, we can approximate the accessible volume in each water `cell' as a regular square bipyramid of side 8.8~nm, Fig.~\ref{Im3m}(b). The centres of lysozyme molecules can explore a smaller bipyramid of side $8.8-\sigma = 5.4$~nm, giving $\alpha_{\rm Im3m} \approx (8.8/5.4)^3 = 4.3$ \cite{imagecharge}. Literature data for MO under different conditions without \cite{Qui00} and with lysozyme \cite{Razumas96} suggest that we can expect variations of up to $\sim \pm 30\%$ in the lattice parameter, so that $\alpha \sim 3-11$. 

In our simulations, we used small spheres to model lysozyme, while 125 larger spheres fixed on a primitive cubic lattice modelled the geometric confinement effect of the MO cubic phase. Such a model has been used before to study fluids in porous media \cite{Page96}.  The small sphere diameter and large-sphere simple cubic lattice constant were taken to be 3.4~nm and 12.5~nm respectively, to agree with the known lysozyme molecular volume and the cubic phase with lysozyme lattice constant at 25$^{\circ}$C \cite{Razumas96}. We simulated a model with a particular $\alpha$ by using a value for the large sphere diameter such that the small-sphere centers can access $\alpha^{-1}$ of the total space between large spheres. Periodic boundary conditions apply.

The small spheres (diameter $\sigma$) interact via a hard core and a square-well attraction:
\begin{equation}
    u(r)=\left\{    \begin{array}{ll}
        +\infty,    &   r<\sigma\\
        -\epsilon   &   \sigma\leq r < (1+\lambda )\sigma\\
        0       &   r\geq(1+\lambda )\sigma\\
            \end{array} \right.,
\end{equation}
where $\lambda$ and $\epsilon$ are the range and depth of the square well in units of $\sigma$ and $k_B T$ respectively. We took $\lambda=0.08$ \cite{Kulkarni03}. Increasing the salt concentration experimentally corresponds to increasing $\epsilon$; in our range of salinity, we expect $0 < \epsilon < 3$ \cite{Kulkarni03}. The small spheres behave as hard spheres to the fixed large spheres. 
 
We investigated the coexistence of a fluid of such attractive small hard spheres confined by a fixed array of large hard spheres with an {\it unconfined} crystal of the attractive hard spheres. This is the thermodynamic situation of crystals nucleating in mesoscopic defects filled with concentrated protein solution `fed' from the cubic phase. Rather than simulating the free energies, we used an approximation to calculate coexistence by simply equating chemical potentials and neglecting the equal-pressure condition (see \cite{Sear99} for a careful justification). 

First, in the spirit of the cell model \cite{Buehler51} we can approximate the volume accessible to the centre of a particle in the crystal to be $\pi\lambda^3\sigma^3/6$, so that the chemical potential of the crystal can be written as, 
\begin{equation}
    \mu_s=\mu_0-n_s\left(\frac{\epsilon}{2}\right)-3k_BT\ln\lambda,
    \label{mu_s}
\end{equation} 
where $n_s$ is the number of particles whose centers lie in the region $\sigma\leq r<(1+\lambda )\sigma$ from a given particle. Here, $\mu_0=k_BT\ln(6\Lambda^3/\pi\sigma^3)$ where $\Lambda$ is the de Broglie (thermal) wavelength. Note that with this choice of $\mu_0$,  the ideal-solution chemical potential is given by $\mu_i=\mu_0+k_BT\ln \phi$.

\begin{figure}
\begin{center}
\rotatebox{-90}{\includegraphics[width=5cm]{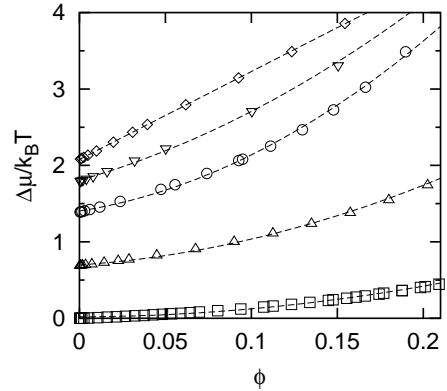}}
\caption{The excess chemical potential for a fluid with $\lambda=0.08$ at $\epsilon=1.5$: $\alpha = 1$ (unconfined, $\square$), 2 ($\triangle$), 4 ($\circ$), 6 ($\triangledown$), and 8 ($\diamond$) from simulation. The lines are guides to the eye.}
\label{mu-phi}
\end{center}
\end{figure}

Next, consider the fluid of confined small spheres. The leading order correction to the ideal limit is
\begin{equation}
    \mu_f   = k_BT\ln(\rho_a\Lambda^3) = \mu_i+k_BT\ln\alpha \;, \label{mu_f}
\end{equation}
where $\rho_a$ is the number density of small spheres in the volume accessible to them, and $\alpha^{-1}$ is the ratio of this accessible volume to the total volume between large spheres. In our simulations, $\alpha = 2.0, 4.0, 6.0$ or 8.0. We used the `Widom insertion method' \cite{widom} to obtain $\Delta \mu = \mu_f - \mu_i$ at different values of $\epsilon$ with and without confinement, Fig.~\ref{mu-phi}. Note that the confined results indeed go to $\Delta \mu/k_B T = \ln \alpha$ in all cases when $\phi \rightarrow 0$.

\begin{figure}[t]
\begin{center}
\rotatebox{-90}{\includegraphics[width=6cm]{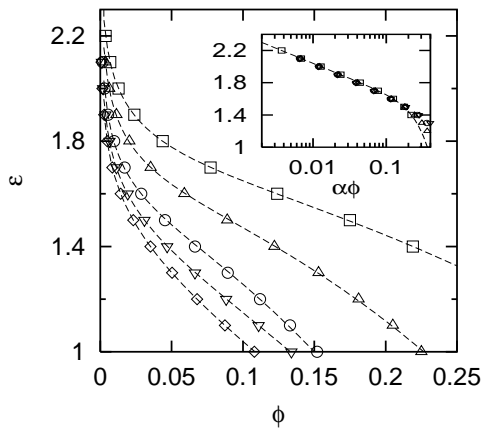}}
\caption{The simulated equilibrium crystallization boundary with $\alpha = 1$ (unconfined, $\square$), 2 ($\triangle$), 4 ($\circ$), 6 ($\triangledown$), and 8 ($\diamond$). The lines are guides to the eye. In the inset, the volume fraction of each system is multiplied by $\alpha$. This superimposes all five boundaries at $\alpha \phi \lesssim 0.2$.}
\label{phase_diagram_sim}
\end{center}
\end{figure}

Now we calculate fluid-crystal coexistence at each value of $\epsilon$ by equating the crystal chemical potential, Equation~\ref{mu_s} with $n_s = 12$, to the simulated fluid chemical potential \cite{Sear99}. The low-density branches with and without confinement are shown in Fig.~\ref{phase_diagram_sim}. Increasingly pronounced confinement indeed systematically decreases the solubility in a way very reminiscent of the data shown in Fig.~\ref{phase_diagram}.

Quantitatively, we show in the inset that multiplying the volume fraction $\phi$ by $\alpha$ leads to overlap of the phase boundaries for $\alpha \phi \lesssim 0.2$. Note that the same result is obtained (data not shown) if we map $\epsilon$ to the salt concentration via measured second virial coefficients \cite{Kulkarni00,Kulkarni03}. 

This result suggests that we should attempt the same procedure on the experimental data. This is problematic. The solubility boundary {\it in cubo} cannot be measured:  Fig.~\ref{phase_diagram} shows a nucleation boundary. If we nevertheless proceed, we find that multiplying the lysozyme concentration by a factor of 4 shifts the confined nucleation boundary to the bulk solubility boundary (inset, Fig.~\ref{phase_diagram}). The {\it in cubo} solubility boundary will undoubtedly occur at lower concentrations than the nucleation boundary. We therefore expect that a `shift factor' of $> 4$ will be needed to translate the (unavailable) {\it in cubo} solubility boundary to overlap with the bulk solubility boundary \cite{caveat}. Given that we expect $\alpha \sim 3-11$ in our cubic phase, this expectation is not inconsistent with our simulation results.

In reporting our experimental findings, we have so far not mentioned any time-dependent features. Two aspects are worthy of note. First, we often observed amorphous precipitates before the appearance of crystals. Unlike crystals, these precipitates would occur rather uniformly throughout a sample, and often appeared as `stars' in the microscope. Sometimes, growing crystals would `consume' surrounding amorphous precipitates. The origin and nature of these precipitates are unclear. Secondly, although it is difficult to be quantitative, it is our impression that the {\it in cubo} growth of crystals is a slower process compared to that in bulk solution. This may, of course, be a reflection of the fact that `feeding' lysozyme into the growth medium residing in defects is a slow process. 

To conclude, we have studied the phase behavior of lysozyme/salt in MO cubic phase. The latter lowers the solubility of the protein. This is predominantly a confinement effect: experiments and simulations suggest that a single `excluded volume scaling factor' can shift the solubility boundary back to its bulk position. The altered kinetics of crystallization {\it in cubo}, relying as we suggest on defects and other properties of the cubic phase, may offer a new means of controlling protein crystal nucleation and growth. The generic, entropic, nature of our proposed explanation suggests that lipid cubic phases may be useful for manipulating a range of nanoparticles, and for studying the basic statistical physics of confinement.

ST held a JSPS Overseas Research Fellowship. We thank Mike Cates for discussions relating to \cite{Sear99} and critical reading of the manuscript.


\begin{references}
\bibitem{imagecharge} In practice, we expect $\alpha > 4.3$, since the charges on the lysozyme lead to repulsive image-charge interactions close to a low-dielectric constant bilayer. At a salt concentration of $\sim 0.2$~M, the protein-bilayer gap has to be considerably smaller than $\sim 0.7$~nm to result in a noticeable effect.
\bibitem{caveat} The bulk nucleation boundary is known \cite{pH4.5phase}, but comparing two kinetic boundaries is just as problematic. If we nevertheless proceed, we find that the factor of 4 gives coincidence for $\phi > 0.1$, while a factor of order 10 is needed for lower $\phi$. This may indicate changes in the MO cubic phase structure and/or the lysozyme nucleation mechanism as protein density changes. But this level of quantitative interpretation is almost certainly unjustified in the spirit of the present work.
\end{references}
\end{document}